\documentclass[AER]{AEA}
\usepackage{tabularx} 
\usepackage[colorlinks=true, linkcolor=blue, citecolor=blue, urlcolor=blue]{hyperref}
\usepackage{xcolor}
\usepackage{graphicx}
\usepackage{amssymb}
\usepackage{amsmath}

\numberwithin{theorem}{section}

\numberwithin{lem}{section}
\newtheorem{definition}{Definition}[section]
\numberwithin{definition}{section}
\newtheorem{ass}{Assumption}[section]
\numberwithin{ass}{section}
\allowdisplaybreaks
\numberwithin{equation}{section}
\usepackage{natbib}
\bibpunct{(}{)}{;}{a}{}{: }
\newcommand{\indep}{\perp \!\!\! \perp}
\usepackage{bbm}
\usepackage{cancel}
\draftSpacing{1.5}

\begin{document}
\title{ Estimating Heterogenous Treatment Effects for Survival Data with Doubly Doubly Robust Estimator}
\shortTitle{Survival HTE}
\author{Guanghui Pan \thanks{Corresponding to Guanghui Pan, Department of Sociology, University of Oxford. 42 Park End Street, Oxford, the United Kingdom, OX1 1HH. \url{guanghui.pan@sociology.ox.ac.uk}. Replication files are available at: \url{https://github.com/pp33pant/survival.}}}
\date{\today}
\maketitle
\section{Causal Inference for Survival Data}\label{sec1}
In the previous introduction chapter, we discussed the general concepts of causal inference and the efficient (doubly robust) estimator for the average treatment effect (ATE). In this chapter, we turn to a very specific extension of the framework to analyze the application of the efficient/debiased/doubly robust casual estimator in survival analysis. \\

Causal inference for survival outcomes is ubiquitous in public health and medical research, as researchers are interested in how the trial changes patients' potential longevity or disease progression. It is also applicable in social science areas like demography and public policy when studies are interested in how a social variable may change the potential life expectancy or existing time in a specific population. \\

In medical studies, researchers may have fully-observed data with \textbf{randomized controlled trials (RCT)}. For example, they trial the effectiveness of a drug on the survival of mice. Under such a scenario, researchers implement the treatment for one of the two randomized groups (making the other group without implementation the control), set the time of the implementation as \(t = T_0\), record the time interval between the start of \(T_0)\) and the demise time for both the treatment and control groups. Then researchers could compare the expectation of the survival time between the treatment and control groups, using statistical methods to make inferences from the sample parameters to the population and derive the causal estimand for the drug. \\

Sometimes, the data structure that the researchers encounter is more complex. On the one hand, researchers will use the observational data for causal inference, which does not specify the assignment of treatment and control with randomization. We have discussed the techniques of causal inference with observational data in the introduction chapter \footnote{Also, as we have discussed in the introductory chapter, the efficient estimators for the average treatment effect given by the difference between the treatment and control expectations are the same for the RCT and the observational data.}. Specifically for the observational survival data, researchers can encounter the missing data scenarios respectively called \textbf{truncation} and \textbf{censoring}. Truncation refers to the situation in which some cases are not included in the dataset because they fall outside of the observational window. \textbf{Censoring} refers to the situation in which the event of demise has not occurred by the end of our observation window, and therefore, the exact time is unknown. We will give a detailed definition of truncation and censoring in Section \ref{sec2}. In short, in the above example exploring the drug's treatment effect on the mice, for observational data, we may observe for some mice the treatment had been implemented before we started our observation, and for some mice, we may never observe their death even our observation comes to an end. \\

Therefore, like the missing counterfactuals in causal inferences, for the survival data containing truncation and censoring, we need an unbiased estimator to estimate the survival time if it has no truncation and censoring \citep{van2003}. This is the doubly-doubly robust estimator for the causal inference on the truncated and censoring survival data we introduce in this chapter: we apply a first doubly robust estimation on the causal effect estimation and a second doubly robust estimation on the survival curve estimation. \\

The organization of this chapter is as follows. Section \ref{sec2} reviews the assumptions we mentioned in the introduction chapter for causal inference and the efficient causal inference estimator. Section \ref{sec3} reviews the notations, basic concepts, and basic assumptions for survival data analysis and machine learning method. We will elaborate on the derivation of the mean survival time and the loss function based on different parametric and nonparametric survival models. Section \ref{sec4} derives the efficient/doubly-robust estimator for the survival analysis.  Section \ref{sec5} summarises the algorithm to apply the doubly-doubly robust estimator to infer the average treatment effects and heterogeneous treatment effects with the survival outcomes. In Section \ref{sec6}, we run simulations comparing our doubly robust estimation model with other model settings. 

\section{Assumptions and Doubly Robust Estimator for Causal Inference}\label{sec2}

This part serves as a very simplified review of our introductory chapter (in separate papers, see \citeauthor{pan2024} \citeyear{pan2024}). From that chapter, we have known that the causal inference with observational data indeed uses the observational estimator for a specific treatment status \(E[Y_i | A_i = a, X_i]\) as the unbiased estimator for the statistical estimand \(E[Y | A = a, X]\) and make inference on the causal estimand of the treatment status \(E[Y(a)]\) (notations here: the observational dataset: \(Z_i = (X_i, A_i, Y_i)\), and the statistical measurable set: \(Z = (Y, A, X)\); \(Y\) denotes the dependent variable/ predictor/ outcome, \(X\) denotes the covariates/independent variable/features, and \(A\) denotes the treatment assignment status), solving the fundamental problem in causal inference that we could not observe \(Y(1)\) and \(Y(0)\) (as the treatment is dichotomous) simultaneously. To make the statistical estimand inferrable on the causal estimand, we have the assumptions for causal inference, which we elaborated on Assumption II.4 in our last chapter:
\begin{ass}[Causal Inference Assumptions]
    Suppose a statistical DGP \(Z = (X, A, Y)\), in which \(X\) denotes the covariates, \(A\) denotes the treatment, and \(Y\) denotes the outcome. To make the statistical estimand \( \psi(\mathcal{P}) = E[E_X[Y | A = 1, X]] - E[E_X[Y | A = 1, X]]\) equivalent to the causal estimand \( \psi^*(\mathcal{P}^*) = E[Y(1)] - E[Y(0)] \) from the causal DGP \(Z^* = (X, A, Y(1), Y(0))\) (where \(Y(1)\), \(Y(0)\) denote the potential outcomes under treatment and control, respectively), we need the following hypotheses: 
    \begin{enumerate}
        \item Positivity: the probability to be assigned to treatment and control group conditioned on the covariates, is a positive number between 0 and 1:
        \[
        P(A = 1 |X) \in (0,1) \quad P(A = 0 | X) \in (0,1)
        \]
        \item Consistency: the potential outcome under the treatment received is the same as the observed outcome: 
        \[
        Y = Y(A)
        \]
        \item Unconfoundedness: conditional on a set of observed covariates \(X\), the potential outcomes \(Y(1)\) and \(Y(0)\) are independent of the treatment assignment \(A\):
        \[
        \{ Y(1), Y(0) \} \indep A | X
        \]
    \end{enumerate}
\end{ass}
Meanwhile, we use the estimator from the observational data to yield an unbiased estimation on the statistical estimand. Under the condition of regularity and asymptotical linearity, with the Cramer-Rao Bound Theorem, we have the unbiased and least variance estimator called efficient/doubly-robust/debiased/Neyman-orthogonal estimator for the average treatment effect (ATE):
\begin{equation}\label{eq:eff_ate}
\hat{\psi}_{\text{DR}} = \frac{1}{n} \sum_{i=1}^n \left[\frac{A_i}{\pi(X_i)} (Y_i - \mu_1(X_i)) - \frac{1 - A_i}{1 - \pi(X_i)} (Y_i - \mu_0(X_i)) + \mu_1(X_i) - \mu_0(X_i) \right].    
\end{equation}
The nuisance functions are \(\pi(X_i) = P(A = 1 |X)\) and \(\mu_a(X_i) = E[Y_i | X_i, A_i = a]\).Equation \ref{eq:eff_ate} estimates the causal estimand for the ATE (\(E[Y(1) - Y(0)]\)) via the statistical estimand (\(E[Y | A= 1, X] - E[Y | A = 0, X]\)). Suppose the outcome variable \(Y_i\) in the observational data indicates the survival (time-to-event) of the individual, this is the data structure we are discussing in this chapter. 

\section{Notations and Basic Concepts of Survival Data Analysis}\label{sec3}
\subsection{Discrete and Continuous Survival Outcomes}
Survival analysis deals with the time-to-event outcome: the death of an animal/ species, the survival time of an unstable atom, and the failure time of a system/ a machine are all time-to-event outcomes. Observing the process of the survival event and making statistical inferences about the population is the key to the survival analysis. Hence, there are two ways to observe the survival outcome: for the first method, and usually for the short-life objectives, we could calculate the accurate \textbf{failure time}, which we treat the survival outcome as the \textbf{continuous} survival outcome; or we could observe whether the event has occurred with specific intervals like hours, days, weeks, and years. We have a dichotomous indicator on whether the object failed, and another indicator showing the number of the interval. We call this the \textbf{discrete survival outcome}.\\

For discrete survival outcomes, the outcome \(Y\) in the statistical estimand can be inferred as the probability of survival grouped by the specific time interval using the frequencies of survived cases in the dataset. Hence, for causal inference, we can identify the causal estimand as \(E[Y_t(1) - Y_t(0)]\) (where \(t \in [0, +\infty)\) and \(t \in \mathbb{Z}^+\); thus \(t\) stands for the specific time interval), which is the difference in the probability of survival for the treatment and control groups (can also be denoted as \(P_t(1) - P_t(0)\), where \(P\) is the probability of failure or survival). Therefore, for every time interval, we may have a doubly robust/efficient estimator on the ATE, and the ATE will change with the change in the time interval. Social scientists applied this data structure to conduct policy analysis. For instance, in \citeauthor{wolfers2006} (\citeyear{wolfers2006}), he initially used a fixed-effect discrete-year setting to study the change of divorce laws on the change in divorce rates. Indeed, the paper measured the probability difference of not-divorced at each time interval between the policy-affected and not-affected groups.\\

Sometimes, in social science, even if the survey is conducted at a fixed interval and we observe the discrete survival outcomes, we will treat the survival outcome as continuous and use continuous survival models to deal with it. This usually requires further assumptions on the data distribution.  In this following chapter, we mainly discuss the estimator for the continuous survival outcomes. The survival outcome, under the continuous scenario, can be written as a function of \(t\). Suppose  \(T\) stands for the time the event of research interests occurs, and we can define the survival function \(S(t)\) on the domain \(t \in [0, T]\) and \(t \in \mathbb{R}\) simply as the survival probability (rate) at time \(t\):
\begin{definition}[Survival Function]\label{def:surv}
    \[
    S(t) = P(T \ge t) 
    \]
    For the conditional survival function, we have:
    \[
    S(t|X) = P(T \ge t \mid X) 
    \]
\end{definition}
Since the cumulative distribution function (CDF) for the random variable \(T\) is defined as: \(F(t \mid X) = P(T \leq t \mid X)\), we could rewrite the conditional survival function \(S(t|X)\) as: 
\[
S(t|X) = 1 - F(t |X) = 1 - \int f(t | X) dt \footnote{As the mathematical transformation of the survival functions is irrelevant with the covariates \(X\), thus, if we are not estimating specific parameters with the covariates, we only discuss the unconditional survival function \(S(t)\), the CDF \(F(t)\), the PDF \(f(t)\), the mean survival time \(E[T]\), and the hazard function \(h(t)\). When we are estimating parameters from the covariates, for instance, when we need the covariates X to estimate the rate parameter \(\lambda\) in exponential distribution, we will give the conditional expression, for example, the PDF \(f(t \mid X)\).}.
\]
The term \(f(t)\) refers to the probability density function of \(t\), which has a relationship with the survival function, deriving from the equation above:
\begin{equation}\label{eq:pdf}
    f(t) = -\frac{d S(t)}{d t}.
\end{equation}

Estimating the survival function to yield the ATE that satisfies the theoretical requirements of the data-generating process is the key to the causal inference with survival outcomes. Obviously, like the discrete survival outcomes, we could still set a specific time point \(t = T\) and compare the survival probability for the treatment group \(S^1(T)\) and the control group \(S^0(T)\) for the ATE \footnote{For the efficient estimators for the ATE in Equation \ref{eq:eff_ate}, indeed, we are using the estimated \(\hat{S}^0(T | X_i)\) and \(\hat{S}^1(T | X_i)\) to simply be \(\hat{\mu}_0(X_i)\) and \(\hat{\mu}_1(X_i)\) for the estimations. We here put the indicator for treatment status \(A =a\) in superscript to differentiate \(S^0(T)\) as the survival function for the control group at time \(T\) and \(S_0(T)\) which is the baseline survival function \(S_0(T) = \exp(-H_0(T))\).}. For instance, \citeauthor{wu2022} (\citeyear{wu2022}) also examined the change of divorce laws on the longevity of marriage. They construct the continuous survival function based on the Cox-Proportional Hazard Assumption (see below), set the observational year as the end of year seven, and applied the difference-in-difference (DID) framework to measure the causal effect \footnote{As mentioned in the introduction chapter, the DID framework solves the violation of the positivity assumption: in the states where the divorce laws were enforced, there were no counterfactual cases in the control group, and the cases in the neighborhood states have to be included in the research as the control. }.\\

Our goal for survival analysis with observational data is to have the survival function estimated from the observational data \(\hat{S}( t \mid A = a, X)\) to be the estimator for the true survival function with specific treatment status \(S^{A = a}(t)\). For the independent and identically distributed (IID) samples, the group-level survival function (for either the treatment or the control group) is the product of all the single survival curves for the individuals in that group, namely, 
\[
\hat{S}(t | A = a, X) = \prod_{i: A_i = a}\hat{S}_i(t |X_i) 
\]
In statistics (and machine learning), we use the \textbf{loss function} to quantify the discrepancy between the observational data and the predictions on the survival function, which takes the form of the negative log-likelihood of the survival function:
\[
\mathfrak{L}:= -\sum_{i=1}\log L(\theta) = -\sum_{i=1}\log(f(t; \theta)).
\]

In which \(theta\) refers to the parameters we estimate. Below, we will introduce several common parametric and nonparametric methods for estimating the survival function using observational data. 

\subsection{Estimating the Complete-Case Loss and Mean Survival Time}
Besides comparing the difference in survival rates, We could also construct the \textbf{mean survival time} separately for the treatment and control groups \footnote{In many studies, researchers traditionally estimate the median survival time, or the half-life during the survival process to represent the survival function. According to the definition, the half-life time is \(T_m: S(T_m) = 1/2\), which is more intuitive and more revealing of the central tendency than the expectation (as the survival time data are usually right-skewed: some cases have longer survival time). However, when we use the median survival time to capture the treatment effect, we could still have the individual treatment effect (ITE) for individuals, but on the group level  (the divergence between the treatment group and the control group), we are capturing the survival median treatment effect (see \citeauthor{hu2021} \citeyear{hu2021}). This is still feasible as long as we can yield the doubly robust/efficient estimator for the median treatment effect (for instance, we could use the heuristic method mentioned in the introduction chapter to accomplish this). However, this is beyond our discussion here as we focus on the treatment effect based on the expectation.}. Usually, for parametric and nonparametric models, without any transformation, they are defined on the domain of \((-\infty, +\infty)\) (for instance, if the outcome is defined on \([0,1]\), we will use sigmoid (logistic) or probit models to transform the original predicted value from the covariates). The time-to-event data (survival time \(T\)) are defined on the domain of \([0, +\infty)\). Therefore, we need appropriate model specifications for the non-negative nature of the survival time and derive the mean (expected) survival time. According to the definition of the expectation, we could derive the relationship between the mean survival time and the survival function as:
\begin{equation}\label{eq:mean_surv}
E[T] = \int_0^\infty t f(t) dt = \int_0^\infty t\left(-\frac{dS(t)}{dt}\right)dt = \int_0 ^\infty S(t) dt. 
\end{equation}
Thus \footnote{Because \(\int_0^\infty t\left(-\frac{dS(t)}{dt}\right)dt = \int_0^\infty -tdS(t) = -[tS(t) \mid_0^\infty - \int_0^\infty S(t)dt] = \int_0^\infty S(t)dt\).}, \(\hat{E}[T |X] = \int_0^\infty \hat{S}(t\mid X)dt\). \\

\subsubsection{Parametric Models and the Hazard Function}
If we apply a parametric model to model the survival data, a prerequisite assumption is we assume that the survival data follow specific types of distribution:
\begin{ass}[Parametric Modeling Assumption]
If we choose to model survival data with the parametric models, we assume that the survival data follow a particular (known) distribution.  
\end{ass}

We introduce three common parametric models for survival data: log-normal, exponential, and Weibull. The log-normal distribution is somehow like the probit modeling for the data on \([0,1]\): which originally spreads over the entire real number line to map it onto a different range. Specifically, for the log-normal distribution, the transformation shifts the range to \([0, +\infty)\). Therefore, the log-normal distribution suggests \(\log(T| X) \sim \mathcal{N}(\mu(X), \sigma(X))\). This model setting is similar to the logistic transformation when the outcome is distributed in \([0, 1]\). The mean survival time if \(\log(T |X)\) is normally distributed is:
\[
E[T | X] = E[\exp(\log(T \mid X))] = \exp\left(E[\log(T|X)] + \frac{\text{Var}(T \mid X)}{2}\right)  = \exp\left(\mu(X) + \frac{\sigma(X)}{2}\right)
\]

If we assume that the time-to-event data is distributed log-normally, we thus have the survival function as:

\[
S(t | X) = P(T > t | X) = 1 - \Phi \left( \frac{\log(t) - \mu(X)}{\sigma(X)} \right)
\]
in which \(\Phi\) refers to the CDF for the standard normal distribution. Thus, the PDF is:
\[
f(t; \mu, \sigma) = \frac{1}{t \sigma \sqrt{2\pi}} e^{-\frac{(\log(t) - \mu)^2}{2\sigma^2}}
\]
The loss function, therefore, is the negative log-likelihood:

\[
\mathfrak{L} = -L(\mu, \sigma) = n \log(\sigma) + \frac{1}{2\sigma^2} \sum_{i=1}^{n} (\log(t_i) - \mu)^2 + \frac{n}{2} \log(2\pi)
\]

Also, in some literature, \(T \mid X\) can be approximated as the \textbf{exponential distribution} with scale parameter \(\lambda\): \( T \mid X \sim \text{Exponential}(\lambda(X))\), where \(\lambda(X)\) is usually the exponential of the linear combination of the covariates \(\mu(X)\). The exponential distribution clearly is defined on the domain of \([0, +\infty)\) and could also address the skewed cases with adjustments on the scale parameter \(\lambda\). According to the definition of the exponential distribution, the survival function is expressed as \(S(t|X) = \exp(-\lambda(X)t) \) \footnote{The probability density function for the exponential distribution is 
\[
f(t) = \lambda \exp(-\lambda t); \quad t \geq 0
\]
Therefore, \(f(t| X) = \lambda(X) \exp(-\lambda(X) t)\). With the relationship between the PDF and the survival function, we have \(f(t|X) = -\frac{dS(t|X)}{dt}\), thus, 
\[
S(t|X) = \int_0^\infty f(t|X) dt = \int_0^\infty \lambda(X) \exp(-\lambda(X) t) dt =  \exp(-\lambda(X) t).
\]}, and the mean for exponential distribution is the inverse of \(\lambda(X)\): 
\[
E[T | X] = \frac{1}{\lambda(X)} = \frac{1}{\exp(\mu(X))}\footnote{Because \(S(t|X) = \exp(-\mu(X)t)\), 
\[
E[T|X] = \int_0^\infty S(t|X) dt =  \int_0^\infty \exp(-\lambda(X)t)dt = \left[-\frac{1}{\lambda(X)}\exp(-\lambda(X)t) \right]_0 ^\infty = \frac{1}{\lambda(X)}.  
\]}
\]

The third parametric model for the survival time usually takes the survival time \(T\) as a \textbf{Weibull distribution}. The Weibull distribution is an extension of the exponential distribution. In the Weibull distribution, we have that scale parameter \(\lambda\) and the shape parameter \(\kappa\), which controls the change of probability of event occurrence over time. The survival function for the Weibull distribution is: \(S(t|X) = \exp\left(-\left(\frac{t}{\lambda(X)}\right)^{\kappa(X)}\right)\). Thus, an exponential distribution is a special Weibull whose \(\kappa = 1\) and \(\lambda\) is the inverse of the scale parameter in the Weibull. To make it more clear why including parameter \(\kappa\) is crucial, we first introduce the concept of the \textbf{hazard function}.\\

As the name indicates, the hazard function indicates the instantaneous possibility (rate) at which the event occurs at time \(t\) (of course, with the condition that the case has survived until time \(t\)). Therefore, the hazard function is defined as a derivative:
\begin{definition}[Hazard Function]
    The hazard function is defined as the limit of the probability that an event occurs in a small time interval, divided by the length of that interval, given that the event has not occurred before time \(t\):
    \begin{equation*}
         h(t) = P(T = t \mid T \geq t)=\lim_{\Delta t \to 0}\frac{P(t \leq T < t + \Delta t \mid T \geq t)}{\Delta t}   
    \end{equation*}

\end{definition}
Based on the definition of the hazard function, we can also derive the relationship between it and the survival function. According to the rule of conditional probability, the nominator of \(h(t)\) can be transformed as: \(P(t \leq T < t + \Delta t \mid T \geq t)= \frac{P(t \leq T < t + \Delta t)}{P( T \geq t)} = \frac{f(t)\Delta t}{S(t)}\). Therefore,
\begin{equation}\label{eq:def_hazard}
h(t) = \lim_{\Delta t \to 0}\frac{\frac{f(t)\Delta t}{S(t)}}{\Delta t} = \frac{f(t)}{S(t)}.    
\end{equation}

We can regard that the survival probability is the opposite \textbf{cumulative hazard}, or regard the hazard as the derivative of the demise (opposite of survival), as the hazard function describes the instantaneous case of demise at the specific time. This can be also shown from the mathematical transformation, combining Equations \ref{eq:pdf} and \ref{eq:def_hazard}, 
\begin{equation}\label{eq:haz_surv}
   h(t) = \frac{f(t)}{S(t)} = \frac{-\frac{dS(t)}{d t}}{S(t)} = -\frac{d S(t)}{S(t)}\frac{1}{d t} = - \frac{d\log S(t)}{d t};
\end{equation}
\begin{equation}\label{eq:surv_haz}
    S(t) =  \exp\left(-\int_0^t h(u)du\right) = \exp\left(-H(t)\right).
\end{equation}

With the relationship between the survival and the hazard function, we now can illustrate that for the Weibull distribution, if \(\kappa = 1\), we indeed assume a constant hazard: the hazard function does not change over time for the exponential distribution since \(h(t | X) = \frac{d \log S(t)}{dt} = \frac{d \cancel{\log} \cancel{\exp}(-\frac{t}{\lambda(X)})}{dt} = -\frac{1}{\lambda(X)}\) (constant over time). Correspondingly, \(\kappa > 1\) indicates that the hazard increases over time, and hence, the event is more likely to occur as time progresses, while \(\kappa < 1\) suggests that the hazard decreases over time. The mean survival time is \footnote{Since \(E[T \mid X] = \int_0^\infty S(t\mid X) dt = \int_0^\infty \exp(-(\frac{t}{\lambda(X)})^{\kappa(X)}) dt \). Let \(u = (\frac{t}{\lambda(X)})^\kappa(X)\), thus, \(t = \lambda(X) u^{1/\kappa(X)}\) and \(dt = \frac{\lambda(X)}{\kappa(X)} u^{(1/\kappa(X)) - 1} du\). Substituting the terms in the integral term, we have \(E[T \mid X] = \int_0^\infty \exp(-u) \cdot \frac{\lambda(X)}{\kappa(X)} u^{(1/\kappa(X)) - 1} du\). The Gamma function \(\Gamma(\cdot)\) is defined as: \(\Gamma(x) = \int_0^\infty u^{x-1} e^{-u} du\). Therefore, we have:
\[
E[T \mid X] = \frac{\lambda(X)}{\kappa(X)} \Gamma\left(\frac{1}{\kappa(X)} + 1\right) = \lambda(X) \Gamma\left(1 + \frac{1}{\kappa(X)}\right).
\]}: 
\[
E[T |X] = \lambda(X) \Gamma\left(1 + \frac{1}{\kappa(X)}\right). 
\]
in which \(\Gamma(\cdot)\) denotes the Gamma function. Usually, we use two sets of functions to fit \(\lambda(X)\) and \(\kappa(X)\) separately. For example, we use the exponential transformation of the two sets of linear combinations of \(X\): \(\lambda(X) = 1/\exp(X\beta)\) and \(\kappa(X) = \exp(X\gamma)\), or nonparametric machine learning models to obtain the results. \\

For the Weibull-class distribution (including exponential), the loss for the survival function is:
\[
\mathfrak{L} = -\sum_{i = 1}\log(f(t_i; \lambda, \kappa))
\]
and as the survival function for the Weibull distribution is \(S(t | X) = P(T > t | X) = e^{-\left(\frac{t}{\lambda(X)}\right)^{\kappa(X)}}\), we have its PDF as:
\[
f(t | X) = 
\frac{\kappa(X)}{\lambda(X)} \left( \frac{t}{\lambda(X)} \right)^{\kappa(X) - 1} e^{-\left(\frac{t}{\lambda(X)}\right)^{\kappa(X)}}
\]
And thus, the empirical loss is given as:
\[
\mathfrak{L} = -L(\lambda, \kappa) = n \log(\lambda) + n \log(\kappa) + (\kappa - 1) \sum_{i=1}^{n} \log(t_i) + \sum_{i=1}^{n} \left(\frac{t_i}{\lambda}\right)^{\kappa}.
\]
\subsubsection{Nonparametric and Semiparametric Model Specifications}
Usually, researchers cannot approximate the distribution of the survival data into any known distribution, or the parametric methods are not applicable. A very common non-parametric modeling for the survival data is called the \textbf{Kaplan-Meier method}, which is quite similar to the discrete method but yields the survival function. Even though many studies on survival data do not apply the Kaplan-Meier method, researchers are used to scratch the \textbf{Kaplan-Meier curve} to show the trend of the survival probability over time. The idea for the Kaplan-Meier method is quite simple: it just counts, stepwise, the number of cases \textbf{at risk} (meaning that they have survived up to the observation time)\(n_j\) and the number of cases the event occurs \(d_j\) at time \(t_j\). So, the survival function from the Kaplan-Meier method is:
\[
S(t) = \prod_{t_j \leq t} \left(1 - \frac{d_j}{n_j}\right)
\]
With the condition that the survival at each time point \(t_j\) is independent. Since we observe the conditions of survival at every \(t_j\), the mean survival time can be approximated by the sum of the survival probability at \(t_j\) times the interval length from the last observation point \(t_{j-1}\) to \(t_j\): 
\[E[T] = \sum_{T_0}^{T_{\max}}S(t_j)(t_j - t_{j-1})
\]
Geometrically, the mean survival time is represented by the square of the area under the survival curve. This applies equally to both parametric models and the Kaplan-Meier model. However, the Kaplan-Meier method does not involve a loss function as the parametric models or the semiparametric Cox Proportional Hazard model to be mentioned below, but here are several loss function analogies with the Kaplan-Meier method to compare the predicted outcome and the true statistical estimand. For instance, we can use the concordance index or the log-rank tests for nonparametric models to analyze the divergence between the Kaplan-Meier estimates and the true statistical estimand. \\

We may also assume some flexible model that does not specify the distribution of the survival data but has some restrictions that could incorporate with the covariates, and even better, it can estimate the effect of a specific covariate on the survival data. This is the most-used model in survival analysis: the \textbf{Cox Proportional-Hazard (Cox-PH)} method, which is, a semiparametric model as it specifies parameters partially in the model. As the name suggests, although it does not assume any distributions for the survival and hazard functions, it restricts the hazard function to be proportional over time across covariates:
\begin{ass}[Proportional Hazard Assumption for Cox-PH Model]
Cox Proportional Hazard models assume the effect of a covariate on the hazard function is multiplicative and does not change with time. Mathematically, the baseline hazard form, \(h_0(t)\) and the conditional hazard \(h(t \mid X)\), both at \(t\), have the following relationship:
\[
h(t \mid X) = h_0(t) \exp(X\beta)
\]
and the term \(\exp(X\beta)\), the \textbf{risk score}, is constant over time.
\end{ass}

For instance, suppose we want to analyze how gender affects survival time. We could derive the hazard ratio between men and women. The assumption here is that the hazard ratio between men and women is constant at any time or interval. For instance, if we suppose men have higher risks than women with the hazard ratio of \(2:1\), then at any given time, the hazard ratio between men and women is \(2:1\). \\

A very convenient way to capture the causal effect of survival data in the previous sociological and demographic studies is to yield the hazard ratio between the treatment and the control group. The estimator is the \textbf{mariginal hazard ratio (MHR)}. However, most of the time, MHR is a biased estimator for causal estimations as it violates the assumption of unconfoundedness (the treatment variable without restricting exogeneity is not independent of the observed outcomes). \\

A circumvent method here is somehow similar to the Inverse Probability Weighting (IPW) or two-step regression: we first predict the propensity for the treatment given the exogenous covariates and use the predicted propensity for the treatment from the first step in the Cox-PH model to yield the coefficient \(\beta\). If the unconfoundedness assumption holds, this will yield an unbiased causal estimand, as the IPW estimator itself is unbiased. However, this method has two shortcomings: first, as we noted in the introduction chapter, the IPW estimator is not an efficient/doubly robust estimator. Thus, the standard error of the estimator will be larger than the most efficient one. More importantly, the causal effect derived from this method is quite hard to interpret. The coefficient suggests the ratio between the hazard functions, which is a relative risk. If we want to show the direction of the treatment (increases or decreases the risks), we could use the estimation of the hazard ratio; however, if we need to interpret quantitatively how much difference the treatment changes for survival, the hazard ratio is not enough.\\

However, the Cox-PH model could be the foundation for us to calculate the survival function \(S^{A = a}(t |X)\) and mean survival time \(E_{A=a}[T | X]\). Cox-PH method provides a toolbox to capture the hazard function \(h( t \mid X\), and using Equation \ref{eq:surv_haz}, we could get the hazard function for \(S( t \mid X)\), and with Equation \ref{eq:mean_surv} we can get the estimated mean survival time. Compared to the estimation of the hazard ratio, if the survival function is required, we need to figure out the baseline hazard \(h_0(t)\). Here, we use the \textbf{Breslow method} \citep{breslow1975} for the estimation on \(\hat{h}_0(t)\) with the observational data. In short, the Breslow estimator suggests that the estimated cumulative baseline hazard \(\hat{H}_0(t)\) is the sum of the inverse of the risk scores for cases at risk at the particular time \(t_j\). Mathematically, it is expressed as :
\[
\hat{H}_0(t) = \sum_{t_j \leq t} \frac{1}{\sum_{i \in R_j} R_i}
\]
Where the risk score, as previously mentioned, if with linear estimation, is the exponentiated linear predictor \(\exp(X_i \hat{\beta})\). Therefore, the baseline hazard function at \(t_j\) is estimated as:
\[
\hat{h}_0(t_j) = \hat{H}_0(t_j) - \hat{H}_0(t_{j-1}) 
\]
Thus, the estimated survival function is:
\[
\hat{S}(t \mid X_i) = \hat{S}_0(t)^{\exp(X_i\hat{\beta})} = \exp\left(-\hat{H}_0(t)\right)^{\exp(X_i\hat{\beta})}.
\]
For Cox-PH models, we usually derive the loss for the hazard function. Since \(h(t \mid X) = h_0(t) \exp(X\beta)\), we may derive the \textbf{partial likelihood function} (ignoring the baseline hazard) to yield the estimation on \(\hat{\beta}\):
\[
\hat{\beta} = \arg \max_\beta L(\beta) = \arg \max_\beta\left[\prod_{t} \frac{\exp( X_i\beta)}{\sum_{j \in R(t_i)} \exp (X_j\beta)} \right]
\]
Where \(R(t_i)\) is the risk set at \(t_i\) (individuals who are still at risk of experiencing the event before \(t_i\) \footnote{ \citeauthor{cox1997}(\citeyear{cox1997}) gives the calculation for partial survival function and its corresponding Hessian matrix for readers who are interested as a reference.}. Thus, the log-likelihood is:
\[
L(\beta) = \sum_{i=1}^n\log\left(\frac{\exp( X_i\beta)}{\sum_{j \in R(t_i)} \exp (X_j\beta)} \right) = \sum_{i=1}^{n} \left[ \beta^T X_i - \log\left(\sum_{j \in R(t_i)} \exp(\beta^T X_j)\right) \right]
\]
As the loss function is the negative log-likelihood, it is:
\[
\mathfrak{L} = -\log L(\beta) = -\sum_{i=1}^{n} \left[ \beta^T X_i - \log\left(\sum_{j \in R(t_i)} \exp(\beta^T X_j)\right) \right]
\] \\

An advantage of the Kaplan-Meier and Cox-PH methods is they are convenient to incorporate with censoring data (see below). For these two models, they calculate the survival function stepwise at each time point, and therefore, for any given time \(t_j\), we can capture the \textbf{restricted mean survival time}, which accumulates the survival from the start point:
\[
\text{RMST}(t_j) =  \int_0^{t_j}S(t)dt. 
\]
\subsection{Machine Learning Framework for Survival Outcomes}\label{subsectionML}
As we derive the loss functions for the survival outcome from both the parametric models (log-normal and Weibull) and the semiparametric Cox-PH model, we can implement the loss in a machine learning algorithm to better reduce the divergence between the predicted survival outcome and the true value. In the doubly doubly robust estimator for the time-to-event data, we use the architecture of the neural network to minimize the loss. For readers who are unfamiliar with the neural network, we briefly introduce its architecture here.\\

In general, the neural network architecture transforms the input data through a series of interconnected layers consisting of multiple neurons applying specific transformations (linear combinations). A basic neural network has three layers: the input layer, which puts the covariates predicting the survival outcome (the features) into the model\footnote{In some literature, the input features are not counted as a layer as they will directly be involved in the mathematical calculations of the hidden layers. Therefore, the first \(K-1\) layers are all hidden layers.}; the hidden layer, which executes specific transformations for the input; and the output layer, which outputs the predicted value and compares it with the observational data. \\

Suppose our neural network has \(K \geq 3\) layers. We start with the input layer. In this layer, every covariate is regarded as a node and takes the form of a vector. The input nodes are transferred into the \(K-2\) hidden layers, and mathematical transformations are executed here. There are two forms of transformations in each layer. The first is a linear transformation. We set \(\Gamma_k\) as the weight/slope parameter of the layer \(k\) and \(\rho_k\) as the bias/intercept parameter. The linear transformation is thus \(X\Gamma_k + \rho_k\). The second transformation is the activation transformation. In order to learn more complex patterns, non-linear functions are applied. For instance, we may use the rectified linear unit (ReLU) function to achieve non-linearity: \(\mathfrak{f}_{\eta_{k:  2 \leq k \leq K-1}} = \max(X\Gamma_k + \rho_k, 0)\) (\(\eta_k = (\Gamma_k^T, \rho_k)^T\) on the \(k-\)th layer). Therefore, after the ReLU transformation, the results are in the range of \([0, +\infty)\). In sum, the hidden layer transformation is indeed \(\hat{F}(X_i) = \prod_{k=1}^{K}\mathfrak{f}_{\eta_{k}}\) (\(\eta =(\eta_1^T, …, \eta_K^T)^T\)). Finally, we feed our result from the hidden layer  \(\hat{F}(X_i)\) into the output layer and with some transformation to compare it with the target outcome to capture the empirical loss. The target outcome could be the hazard or survival rate without functional form, the mean survival time, or the parameters set in our parametric or semiparametric models. In other words, we could directly use neural networks to predict the survival time \(\hat{T}\)and compare it with observed survival time \(T_i\), or we could predict the parameters of the Cox-PH model with the neural network \(\hat{\beta}\), optimize it and use the Cox-PH transformation discussed in the last subsection to generate the mean survival time. In this part, researchers need to pay attention to the domain of the data. For instance, for the parameters in the CoX-PH model \(\beta\), as it can be valued in the whole real number set, we directly use linear activation in the output layer, and it could produce any real number. Likewise, if our outcome is the survival rate in the range of \([0,1]\), we may use sigmoid transformation in the output layer. If the outcome is the hazard function and the survival time is defined on \([0, +\infty)\), we can use the exponential transformation so that our predicted outcome is in that range.\\

The process described above (from the input through the hidden layer to the output layer) is called \textbf{forward propagation}. After calculating the empirical loss between the predicted outcomes from the output layer and the target outcome, we further perform the \textbf{backward propogation} to update every parameter in the model \(\eta_k\) in order to get the minimized loss. To achieve this, we compute the gradient of the loss function with respect to each parameter using the chain rule:
\[
\Gamma_k \leftarrow \Gamma_k - \iota \frac{\partial \mathfrak{L}}{\partial \Gamma_k}; \quad \rho_k \leftarrow \rho_k - \iota \frac{\partial \mathfrak{L}}{\partial \rho_k};
\]
We set parameter \(\iota\) as the \textbf{learning rate}. With multiple training iterations, the neural network could minimize the loss and adjust the parameters to predict the outcome better. Optimization algorithms like stochastic gradient descent (SGD) or Adam are deployed in this process. \\

As mentioned above, we could set the outcome in the loss function directly for the survival time, or the parameters in our parametric and semiparametric models. If we directly set the survival time as the outcome, we call the model a deep survival learning model or simply a neural network (NN) model for survival outcome. If our predictors are the parameters for log-normal survival distribution, we call the model NN-log-normal model. Correspondingly, we have the NN-Exponential, NN-Weibull, and NN-Cox-PH models for the predicted outcomes, respectively, as the parameters for the exponential, Weibull distributions, and Cox-PH models. After obtaining the corresponding survival or hazard function, we could transfer them to capture the mean survival time and apply it to our doubly robust/efficient estimator for the ATE to get the ATE for the survival outcomes. \\

However, in social science and medical research, the observed survival outcome usually contains missing data, and thus, we could not have the complete loss function in our modeling. Dealing with the missing data for the survival outcome is a crucial subject in our method. 

\subsection{Truncation and Censoring}
In general, survival data are missing because the time-to-event is not in our observation window. In some cases, if they are excluded from the analysis because of the occurrence of the event relative to our observational window, this type of missing is called \textbf{truncation}. In some cases, we could only observe their survival status partially, but the time they experience the event is beyond our observation window, and this missing is \textbf{censoring}. More specifically, if individuals are excluded from the analysis as they experienced the event before the time point, for instance, if we start to observe after one year of the treatment and some individuals have already died and not in our samples, we call the cases \textbf{right truncated}; and if cases are observed years after the treatment starts, for instance, we observe the cases one year after the treatment, the cases were \textbf{left truncated}. Similarly, for censoring, if the individuals experienced the time of the event after our observation ends, it is \textbf{right censoring}; and if the individual has experienced the event before the start of our observation but the exact time is unknown, we consider the problem as \textbf{left censoring}. Furthermore, if we observe the event periodically, while the event occurs within a certain time interval but the exact time is unknown, we call it \textbf{interval censoring}. \\

For data in social science and demography, as we use \textbf{coarsening data}, we do not consider the issue of interval censoring as if the event occurs in our observational window; by default, we will approximate its survival time as we first observe the event. We are mainly concerned with the missing left truncation and right censoring. For left truncation, if we conduct observations after a period of time following the start of the experiment, it means that the group-level survival rate at our observation start time is lower than \(1\) (in contrast, without left-truncation the start survival rate for both the treatment and the control groups are \(1\)). For right-censoring, as we are unaware of when the real time-to-event is, our estimation based on the complete survival data will doubtlessly be biased. The method in this paper mainly deals with the \textbf{left-truncated-right-censored (LTRC)} survival data. LTRC data are so common in social science research. For instance, the empirical case in the next chapter aims to explore how widowhood affects the survival outcome of mortality (as a survival outcome, life expectancy) among the elders. We start our observation age at 50 so that individuals widowed before 45 are left-truncated, and individuals who are still alive in our last survey (observation) are right-censoring. \\

Like other missing data analyses, assumptions are required for missing data patterns. In this paper, for the LTRC data, we assume that truncation is \textbf{missing at random (MAR)}, and censoring is also MAR, after controlling the covariates:
\begin{ass}[Assumption for Truncation and Censoring]\label{ass:miss}
    Suppose \(\tau > 0\) denotes the time of truncation, \(C_i > 0\) denotes the censoring time, \(T_i\) is the failure/survival time, \(\delta\) is the indicator for censoring: \(\delta_i = 1 \iff T_i < C_i\) (\(\delta = 0\) means censoring and \(\delta = 1\) means observed), we assume the following conditions are satisfied: 
    \begin{itemize}
        \item The truncation time is independent of the survival time:
        \[
        \tau \indep T_i \iff P(\tau > t \mid T_i) = P( \tau >t)
        \]
        \item The censoring time is independent of the survival time, given the observed covariates:
        \[
        C_i \indep T_i \mid X_i \iff P(C_i > t \mid T_i, X_i) = P(C_i >t \mid X_i)
        \]
        \item The truncation and censoring mechanisms are conditionally independent of each other given the covariates:
        \[
         C_i \indep \tau \iff P(\tau, C_i \mid T_i, X_i) = P(\tau) \cdot P(C_i \mid X_i)
        \]
    \end{itemize}
\end{ass}
With the independence of truncation and censoring, we may further denote \(\delta_i(\tau) = 1 \iff (T_i > \tau) \vee (T_i < C_i)\)\footnote{We could further infer that \(T_i < C_i \iff \delta_iT_i \leq \tau\).}, which is the complete observed cases. Obviously, the set \(\delta_i(\tau) = 1\) is a subset for \(\delta_i = 1\). Like for the survival function \(S(t \mid X_i) = P(T_i > t \mid X_i)\) for the prediction of the survival time, we have the censoring function \(G(t \mid X_i) = P(C_i > t \mid X_i)\) to predict the time for censoring. Correspondingly, we could have the hazard function for censoring, denoted as \(h_G(t \mid X_i) = P(C_i = t \mid C_i \geq t, T_i \geq t, X_i) = \frac{d\log G(t \mid X)}{dt}\) and the cumulative hazard \(H_G(t\ mid X) = -\log G(t\mid X)\) Based on Assumption \ref{ass:miss}, the survival function and the censoring function are independent. \\

Due to the missing values on time-to-event, the loss function served in the models is no longer obvious, as we couldn't obtain the likelihood function directly from the dataset. Therefore, we must apply efficient/doubly robust estimation techniques to the empirical loss so that we can yield the estimator for the mean survival time and apply it once again to the efficient/doubly robust estimator for the treatment effect. As in the process, we used the technique of efficient/doubly robust estimator twice, once for the loss function estimation of the survival outcome and once for the causal effect estimation. Therefore, we call our method a doubly doubly robust estimator. 

\section{Doubly Robust Loss for the LTRC Survival Outcomes} \label{sec4}
In this section, we will elaborate on how we derive the doubly robust/efficient loss for the LTRC survival outcome. Since truncation and censoring are independent processes, and truncation does not affect the relative distribution for survival time, we start with the scenario that only contains the censoring case. \\

Like our process in the introduction chapter for the efficient estimator on non-saturated models, we begin our estimation from a regular and asymptotically linear (RAL) estimator. The IPW estimator of the survival loss is a RAL estimator, with the form of the inverse of the censoring probability. Let \(\tilde{T}_i\) denote the observed time \(\tilde{T}_i = min(T_i, C_i)\) and \(\mathfrak{L}(\hat{F}(X_i), \theta)\) denotes the loss obtained from the complete data for the neural network in Subsection \ref{subsectionML} between the predicted value \(\hat{F}(X_i)\) for the target \(\theta\) (as noted above, could be the survival time, the survival function, the hazard function, etc.), we have:
\begin{equation}\label{eq:ipw}
\mathfrak{L}^{IPW} = \frac{1}{n}\sum_{i = 1}^n\frac{\delta_i}{\hat{G}(\tilde{T}_i \mid X_i)}\mathfrak{L}(\hat{F}(X_i), \theta)    
\end{equation}

As by definition \(G(\tilde{T}_i \mid X_i) > 0\). Since censoring is MAR given the covariates, the estimator from Equation \ref{eq:ipw} should have the same expectation as  \(\frac{1}{n}\sum_{i=1}^N \mathfrak{L}(\hat{F}(X_i), \theta)\). \\

In the introductory chapter, we have shown the efficient influence function (EIF) for the unconditioned expectation \(\psi = E[Y]\) is \(\phi^\dagger = Y - E[Y]\), and the EIF for conditional expectation \(\psi = E[Y | X = x]\) is  \(\phi^\dagger = \frac{\mathbbm{1}(X =x)}{P(X = x)}(Y - E[Y \mid X]) + (E[Y \mid X] - E[Y])\). Therefore, for the fully-observed data (without censoring and truncation), the EIF for the loss function on \(\theta\) is:
\[
\phi^\dagger_{\text{Full-data}}= \mathfrak{L}(\hat{F}(X_i), \theta) - E[\mathfrak{L}(\hat{F}(X_i), \theta)]
\]
Now, we take censoring into consideration. Notice Equation \ref{eq:ipw} has almost the same structure as we derive the EIF for the conditioned treatment effect \(\phi^\dagger(\psi(a)) = \phi^\dagger(E[E_X[Y | A= a, X]] = \frac{\mathbbm{1}(A = a)}{P(A=a \mid X)}[Y - E[Y \mid A = a, X]] + (E[Y \mid A = a, X] - E[Y \mid A = a])\) in our causal analysis (Equation III.23 in the introduction chapter), Thus, we could similarly derive the EIF and the efficient estimator for the loss as:
\begin{align}
\phi^\dagger_{\text{Censoring}} &= \frac{\delta}{G(\tilde{T}\mid X)}\left(\mathfrak{L}(\hat{F}(X), \theta) - E[\mathfrak{L}(\hat{F}(X), \theta) \mid \tilde{T} \geq t]\right) + E[\mathfrak{L}(\hat{F}(X), \theta) \mid \tilde{T} \geq t] - E[\mathfrak{L}(\hat{F}(X), \theta)];  
\end{align}
\begin{align}\label{eq:AIPW}
\mathfrak{L}^{DR}_{\text{Censoring}} &= \frac{1}{n}\sum_{i=1}^n\left[\frac{\delta_i}{\hat{G}(\tilde{T_i}\mid X_i)}\left(\mathfrak{L}(\hat{F}(X_i), \theta) - \hat{E}[\mathfrak{L}(\hat{F}(X), \theta) \mid \tilde{T} \geq t]\right) + \hat{E}[\mathfrak{L}(\hat{F}(X), \theta) \mid \tilde{T} \geq t]\right] \\ &= \frac{1}{n}\sum_{i=1}^n\left[\underbrace{\frac{\delta_i \mathfrak{L}(\hat{F}(X_i), \theta)}{\hat{G}(\tilde{T_i}\mid X_i)}}_{: a} + \underbrace{\left(1 - \frac{\delta_i}{\hat{G}(\tilde{T}_i \mid X_i)}\right)\hat{E}[\mathfrak{L}(\hat{F}(X), \theta) \mid \tilde{T} \geq t]}_{:b} \right] \notag
\end{align}
In survival analysis literature, Equation \ref{eq:AIPW} is usually called the \textbf{augmented inverse probability weighted complete-case (AIPWCC)} estimator (\citeauthor{tsiatis2006} \citeyear{tsiatis2006}, chapter 9, pp. 199-220), as part \(a\) is the IPW for the complete case loss and part \(b\) is the augmented term which more efficiently uses information from the censoring individuals. Let \(\hat{U}(X_i, t) = \hat{E}[\mathfrak{L}(\hat{F}(X_i), \theta) \mid \tilde{T} \geq t]\), the estimation will be doubly robust either we correctly specified \(\hat{G}(t \mid X_i)\) or \(\hat{U}(X_i, t)\). \\

However, Equation \ref{eq:AIPW} is uncommonly seen in the survival literature \footnote{This is because the way we derive the doubly robust estimator for the survival loss is not strictly mathematical derivation. Instead, for readers with social science and demographic backgrounds, we simplified the derivation of the doubly robust estimator by analogy with deriving the conditioned treatment effect, and then we will prove that the derived doubly robust estimator for the survival loss is intrinsically the same as the results from the survival analysis literature. For the rigorous proof, see textbooks like \citeauthor{bickel1993}(\citeyear{bickel1993}) and \citeauthor{tsiatis2006} (\citeyear{tsiatis2006}).}. In most literature, they will simplify part \(b\) as:
\[
\left(1 - \frac{\delta_i}{\hat{G}(\tilde{T}_i \mid X_i)}\right)\hat{U}(X_i, t) = \int_0^\infty \frac{\hat{U}(X_i,t)}{\hat{G}(t \mid X_i)} dM_G(t \mid X_i)
\]
As function \(M_G(t \mid X)\) is the censoring martingale at \(t\) given the covariates \(X\). To understand this, we first introduce the definition of the martingale: 
\begin{definition}
Martingale is a process describing the difference between a counting process (\(N(t)\)) and the "compensator" generated by the intensity function (\(\Lambda(t) = \int_0^t h(u)du\)):
\[
M(t) = N(t) - \int_0^t h(u)du
\]
\end{definition}
Let's say we have counted the number of items censoring at time \(t\): \(N(t) = \mathbbm{1}_{\delta = 0, \tilde{T} \leq t}\) (the actual number of case censoring from the observational data). Besides, we could also calculate the predicted counts for censoring from the model relevant to censoring, for example, with the censoring hazard function: \(\Lambda(t) = \int_0^t \mathbbm{1}_{\tilde{T}\geq u} h_G( u \mid X)\). Therefore, the \textbf{censoring martingale} represents the cumulative excessive amount of censoring from time \(0\) to \(t\):
\[
M_G( t \mid X) = \mathbbm{1}_{\delta = 0, \tilde{T} \leq t} - \int_0^t \mathbbm{1}_{\tilde{T}\geq u} h_G( u \mid X)
\]
With the definition, we may infer that \citep{strawderman2000,robins1992}:
\[
1 - \frac{\delta_i}{G(\tilde{T} \mid X)} = \int_0^\infty \frac{d M_G(t \mid X)}{G(t \mid X)}
\]

Simply because 
\(\int_0^\infty d M_G(t \mid X) = (1-\delta_i) - (1- G(\tilde{T} \mid X) ) = G(\tilde{T} \mid X) - \delta_i\). Therefore, we may rewrite Equation \ref{eq:AIPW} into the form with the martingale:
\begin{align}
\mathfrak{L}^{DR} &= \frac{1}{n}\sum_{i=1}^n \left[\frac{\delta_i \mathfrak{L}(\hat{F}(X_i), \theta)}{\hat{G}(\tilde{T_i}\mid X_i)} + \int_0^\infty \frac{\hat{U}(X_i,t)}{\hat{G}(t \mid X_i)} dM_G(t \mid X_i)\right] \label{eq:DR1}\\ &=   \frac{1}{n}\sum_{i=1}^n \left[\frac{\delta_i \mathfrak{L}(\hat{F}(X_i), \theta)}{\hat{G}(\tilde{T_i}\mid X_i)} + \left(\frac{(1-\delta_i)\hat{U}(\tilde{T}_i, X_i)}{\hat{G}(\tilde{T}_i, X_i)} - \int_0^{\tilde{T}_i}\frac{\hat{U}(t, X_i)}{\hat{G}(t \mid X_i)}dH_G(t \mid X_i)\right)\right] \label{eq:DR2}
\end{align}
While Equation \ref{eq:DR2} just expands Equation \ref{eq:DR1} with the definition of martingale.\\

Finally, we consider the restrictions of left truncation. Since the truncation process is independent of the censoring process, we may just rewrite the expressions in  Equation \ref{eq:DR1} and  Equation \ref{eq:DR2}, changing \(\delta_i\) into \(\delta_i(\tau)\) and \(\tilde{T}_i\) into \(\tilde{T}_i(\tau)\) (indicating that we observe partially of the survival after the truncation time \(\tau\)):
\begin{align}
          \hat{\mathfrak{L}}^{DR}_{LTRC} &= \frac{1}{n}\sum_{i=1}^N\left[\frac{\delta_i(\tau)\mathfrak{L}(\hat{F}(X_i), \theta)}{\hat{G}(\tilde{T_i}\mid X_i)} + \left(1 - \frac{\delta_i(\tau)}{\hat{G}(\tilde{T}_i \mid X_i)}\right)\hat{U}(X_i, \tilde{T}(\tau))\right]\label{eq:DR_LTRC1}\\ &=\frac{1}{n}\sum_{i=1}^N\bigg[ \frac{\delta_i(\tau) \mathfrak{L}(\hat{F}(X_i),\theta)}{\widehat{G}(\Tilde{T_i}(\tau)|X_i)} +\int_0^\infty \frac{\widehat{U}(t,X_i)}{\widehat{G}(t|X_i)}d\hat{M_G}(t|X_i)\bigg] \label{eq:DR_LTRC2}\\&= \frac{1}{n} \sum_{i=1}^n  \bigg[\frac{\delta_i(\tau)\mathfrak{L}(\hat{F}(X_i),\theta)}{\widehat{G}(\Tilde{T_i}(\tau)|X_i)}  +  \bigg(\frac{(1-\delta_i(\tau))\widehat{U}(\Tilde{T_i}(\tau),X_i)}{\widehat{G}(\Tilde{T_i}|X_i)} - \int_0^{\Tilde{T_i}(\tau)}\frac{\widehat{U}(t,X_i)}{\widehat{G}(t|X_i)} dH_G(t|X_i) \bigg)\bigg] \label{eq:DR_LTRC3}
\end{align}
In summary, the procedures to apply the doubly robust estimator for the LTRC survival outcomes are as follows: first, based on the observations, we estimate the survival of censoring \(\widehat{G}(\tilde{T}_i(\tau) \mid X_i)\), and construct the architecture of the neural network for the complete case to yield the loss function of the target (survival function, hazard function, mean survival time, log-normal/exponential/Weibull/Cox-PH model parameters) under \(T(\tau) \geq t\): \(\hat{U}(X_i,\tilde{T}(\tau))\); then we use the doubly robust characteristics for the loss function in either Equation \ref{eq:DR_LTRC1}, \ref{eq:DR_LTRC2}, or \ref{eq:DR_LTRC3} to solve the parameters (\(\theta = \arg \min_{\theta} \mathfrak{L}^{DR}_{LTRC}\)); and finally, based on the model we trained, we predict the mean survival time and apply it to our causal inference framework. \\

Sometimes, researchers might find it hard to solve \(\theta = \arg \min_{\theta} \mathfrak{L}^{DR}_{LTRC}\) even with iterations. In such cases, approximate methods may be required. Optimal restriction assumptions may be further needed. For the technical details, see \citeauthor{tsiatis2006} (\citeyear{tsiatis2006}: Chapter 12).   

\section{Doubly Doubly Robust Estimation Algorithm}\label{sec5}
In sum, based on the two doubly robust estimators in Section \ref{sec2} and \ref{sec4}, we develop a doubly doubly robust estimator to capture the average treatment effects and the heterogeneity treatment effect in left-truncated-right-censored survival outcome: we use the first doubly robust estimator to estimate the loss function for the survival outcome and yield the estimation for mean survival time; we then apply a doubly robust estimator to have the efficient/doubly robust estimator for the causal outcome. \\

Beyond that, we can capture the \textbf{heterogeneous treatment effect (HTE)} for the ATE over the spectrum of particular covariates. As it is conditioned on the particular \(X = x\) for the ATE, it is called the conditioned average treatment effect (CATE):
\[
CATE = E[ATE \mid X =x]
\]

Therefore, we can have the procedure to estimate the CATE with survival outcomes as follows: 
\begin{itemize}
    \item First, identify our measurable set \( D_i = {X_i, T_i, Z_i, A_i, C_i, R_i, \tau, \delta_i}\). In the dataset, \(X_i\) denotes the covariate for the CATE, \(T_i\) denotes the observable survival time, \(Z_i\) are the covariates controlling randomization, and \(R_i\) are the covariates controlling censoring, \(A_i\) is the treatment assignment, \(C_i\) denotes the censoring time, \(\tau\) denotes the truncation time, \(\delta_i\) is the censoring indicator indicating the case is still observable if \(delta_i = 1\). Further, based on the variables that appeared in the dataset, we can generate \(\tilde{T}_i(\tau) = (min (T_i, C_i), \wedge \tau\), and \(\delta_i(\tau) = \mathbbm{\tilde{T}_i > \tau} + \delta_i\mathbbm{1}(\tilde{T}_i \geq \tau)\). Then, we divide the dataset \(D\) into three subsets \(D_1, D_2\), and \(D_3\). 
    \item We use \(D_1\) to fit the nuisance parameter \(\pi\) in causal inference. Namely, we have \(\hat{\pi}(Z_1)\): \(\hat{\pi}(Z_1) = P(A_{1i} = 1 | Z_{1i})\) from \(D_1\).
    \item We use \(D_2\) to fit the nuisance parameter \(\mu_a\) in causal inference. As we noted above, in our study, \(\mu_a\) is the mean survival time under the treatment status \(A = a\). 
    \item Regress the pseudo-outcome using the third dataset and yield the doubly robust estimation for the conditional average treatment effects: \(\widehat{CATE}^{DR} =  E\bigg[\big[\frac{A_{3i} - \hat{\pi}}{\hat{\pi}(1-\hat{\pi})} (Y - \widehat{\mu}_{A_{3i}})+ (\widehat{\mu_1} - \widehat{\mu_0})\big] | X = X_{3i}\bigg]\). 
    \item Rotate the three datasets in the previous steps for cross-validation. In other words, separately use \(D_2\) and \(D_3\) to fit the nuisance function for propensity; use \(D_3\) and \(D_1\) to fit the nuisance function for survival function, and use \(D_1\) and \(D_2\) to generate the doubly robust estimation for the CATE. Finally, average the three results. 
\end{itemize}

The algorithm is indeed exactly the same as the algorithm for the efficient/doubly robust/debiased ATE/ CATE. The specific part that needs to be addressed is in Step 3, in which we generate the mean survival time with neural networks with the doubly robust loss:
\begin{itemize}
    \item Split \(D_2\) into \(M\) subsets: \(D_2^{(1)},...,D_2^{(M)}\). 
    \item For each subset \(D_2^{(m)}\), decide the target parameter for the loss function. The target parameter can be the survival rate, instantaneous or cumulative hazard, mean survival time, and log-normal/exponential/Weibull/Cox-PH model parameters. calculate the loss function from the complete case \(\mathfrak{L}(\hat{F}(R_i), \theta)\) and its conditioned expectation \(\hat{U}(R_i, \tilde{T}_i(\tau) = E[\mathfrak{L}(\hat{F}(R_i), \theta) \mid \tilde{T}_i(\tau) \geq t\). Calculate the survival function for censoring \(\hat{G}(\tilde{T}_i(\tau) \mid R_i) = P(\tilde{T}_i\tau \geq C_i \mid R_i). \)
    \item Derive the empirical average estimated loss function based on either Equation \ref{eq:DR_LTRC1}, \ref{eq:DR_LTRC2}, or \ref{eq:DR_LTRC3}. Predict the parameter \(\theta = \sum_{m=1}^M \eta_m \mathbbm{1}_{R\in D_2^{(m)}}\).
    \item Based on the optimized parameter from the last step, correspondingly derive the correct estimation for the mean survival time.  
\end{itemize}
\section{Simulation Work} \label{sec6}
\subsection{Model Settings}
This section uses a simulation to illustrate our method. In total, we examine the estimation from five models: first, we estimate the ATE and the CATE without an RCT framework: we use the effects via the marginal hazard ratio (MHR) directly from the Cox Proportional Hazard Model with specific baseline settings. Then we turn to the RCT settings. We initially use the separated Cox-PH models to fit the survival functions for the treatment and the control and then predict the expected outcome under the treatment and control for the whole dataset with the treatment and control survival functions. Then we use a neural network with a doubly robust loss function to fit the treatment and survival function and to calculate the average and heterogeneous treatment effects in the same way as the Cox-PH models. Since the two models give out naïve plug-in estimations for the treatment effects (for both the ATE and the HTE), we call the naïve Cox-PH model and the naïve “NN-DR loss” models separately. Finally, we apply the Cox-PH and NN-DR loss models under the doubly robust/debased treatment effects framework. We call them the debiased Cox-PH and debiased NN-DR loss models separately. The debiased NN-DR loss model yields the doubly-doubly robust estimator for the average and heterogeneous treatment effects for the survival data. The target parameter in the loss function we choose is the parameter from the Cox-PH method \(\beta\). \\
\begin{figure} 
\includegraphics[width = 1\textwidth]{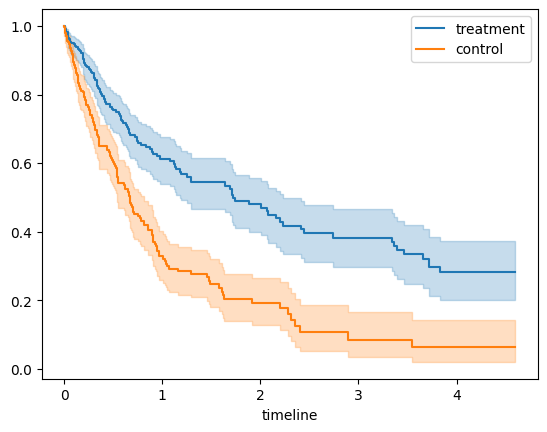}
\caption{Kaplan-Meier Survival Curves for the Simulated Treatment and Control groups}\label{fig1}
\end{figure}

The settings of our simulated data are as follows. We use 500 independent observations. We have 10 covariates for the survival function, \(X_1 - X_10\), uniformly distributed on integers 1 – 100. \(Z_1, Z_2\) are uniformly distributed from 0 to 1. We label cases for which both \(Z_1 > 1/\sqrt{2}\) and \(Z_2 > 1/\sqrt{2}\) as treated, and others are grouped into control. Furthermore, We suppose the failure times are exponential with the mean parameter \(a\) equal to 2 for the treated and 1 for the control, and based on the observed marginal failure time, we set truncation time at the 95th percentile of the distribution (which means 5\% of the cases are truncated). Finally, we set the censoring variable with an exponential distribution with a specific rate parameter of 1.5. Due to our model settings, the simulated data follow the proportional hazard assumption. Moreover, we should expect there to be no statistically significant correlation between the covariates and the survival times and there’s no statistically significant heterogeneity along the propensity score—in other words, the heterogeneous treatment effect line should be paralleled with the average treatment line and horizontal on the span of the propensity scores. \\

With our randomization settings and the simulation specifications, in the simulated data, there are a total of 266 cases allocated to the treatment group, while 234 cases in the control group. 25 cases are left-truncated, and 261 cases are right-censored. The median survival time for the treatment group is 1.86, while the median survival time for the control group is 1.00. Thus, if we use the median survival time between the treatment group and the control group for the average treatment effect estimation, the value should be 0.86. Figure \ref{fig1} below shows the Kaplan-Meier survival curves separately for the treatment and the control groups: \\

Meanwhile, we use the logistic regression method to predict the propensity scores for the treatment with \(Z_1\) and \(Z_2\) as the covariates. The propensity score distribution for the treatment and the control groups can be seen in Figure 2. 
\begin{figure} 
\includegraphics[width = 1\textwidth]{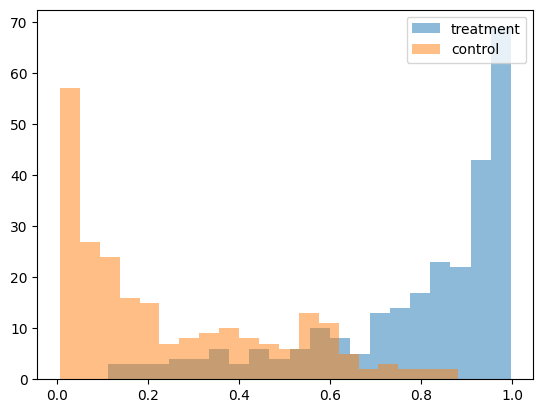}
\caption{Predicted Propensity for treatment and control groups}\label{fig2}
\begin{figurenotes}
    Predicted with logistic regression. 
\end{figurenotes}
\end{figure}
\subsection{Results from non-RCT Models}
\begin{figure} 
\includegraphics[width = 1\textwidth]{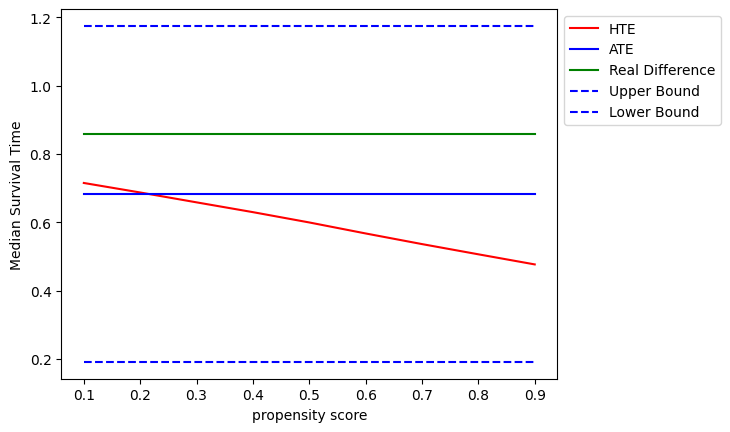}
\caption{ATE and HTE from MHR Model}\label{fig3}

\end{figure}

We first report the results via the marginal hazard ratio (MHR) directly from the Cox proportional hazard (Cox-PH) model. To make it easier for the readers to understand and consistent with our previous sections' illustrations, we demonstrate the ATE and the HTE. The model specification is quite simple: for all individuals, we just fit them with one Cox-PH model. Then we use the coefficient of the variable “treatment” in the Cox-PH model combined with the baseline hazard model to construct the ATE. We assume that Assumptions 1-5 are not violated (especially the proportional hazard assumption). Then we fix the propensity scores at different levels and calculate the HTE. The standard deviation is yielded directly from the model. Figure \ref{fig3} shows the model's ATE and HTE estimation. As can be seen from the figure, the estimation of the ATE is 0.63, with a confidence interval between 0.20 and 1.20. The HTE, meanwhile, has a clear downward trend as the propensity score increases: the estimation of the treatment effect shrinks from around 0.71 at the lowest decile to around 0.47 at the highest decile propensity score. Therefore, although the estimation from the MHR is the most intuitive one, it does not well estimate the ATE and the HTE, and it does not take the selectivity of treatment and control into account. \\

\subsection{Results from Naïve Plug-in Estimators}
Then, we turn to the results from the naïve plug-in estimators. We begin with the simple Cox-PH model setting. We first split the data with the treatment status and separately fit models for data in the treatment and control groups. The mathematical idea for this model can be seen in Section \ref{subsectionML} of this chapter. We have one Cox-PH model from the treatment, and we call it the treatment model. We use the treatment model to predict the outcome for the whole dataset and generate the expected outcome under treatment (a counterfactual outcome). The same process is taken for the control group, and we have the expected outcome under control. Then we substitute the two expected values and derive the individual treatment effects. The average treatment effects from the naïve Cox-PH model take the mean of the individual treatment effects. \\

\begin{figure} 
\includegraphics[width = 1\textwidth]{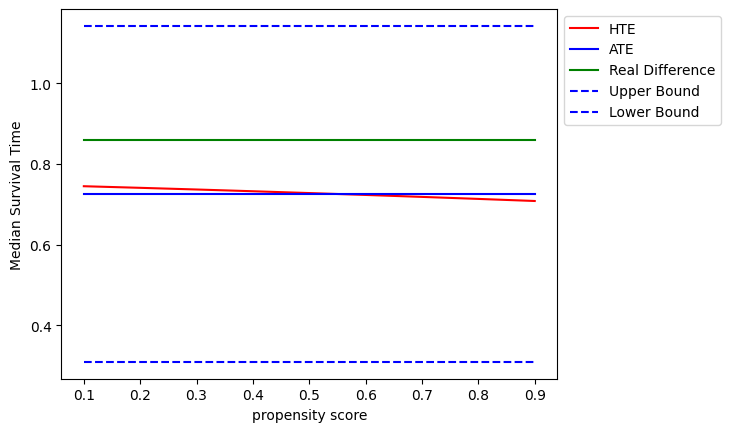}
\caption{ATE and HTE from Naive Plug-in Cox-PH model}\label{fig4}
\begin{figurenotes}
    The standard deviation is bootstrapped from 100 loops. 
\end{figurenotes}
\end{figure}

Furthermore, we resample the whole dataset and bootstrap the procedures above to capture the standard deviation of the ATE. We then fix the propensity score at every decile in our prediction step to derive the HTE. The results from the naïve plug-in Cox-PH models can be seen in Figure \ref{fig4}. The point estimation for the ATE is 0.73, with a bootstrapped standard deviation of 0.21. A slight downward trend in the HTE can also be seen from the graph. However, the slope for the HTE line is smoother than that from the non-RCT model, with an estimation of 0.75 for the first decile and 0.71 for the last decile. \\

We next turn to the naïve plug-in model with the neural network structure and a doubly robust loss. We discussed the mathematical derivation for the architectures of the neural network and the construction of the loss function in Section \ref{sec4} of this chapter. In our model, the neural network has four layers: one input, two hidden, and one output layer. We set the hidden layers as fully connected with 32 nodes, applying a linear transformation on the result from the last layer followed by a rectified linear unit (ReLU) activation function. For the output layer, we set it as a fully connected layer with a single node followed by the Sigmoid activation function. The loss function is the doubly robust survival loss, and we choose the Adam optimizer to adjust the weights based on the loss function. We train the model with 100 epochs, and for each epoch, we set the batch size for both the training and the validation data as 256.  \\

\begin{figure} 
\includegraphics[width = 1\textwidth]{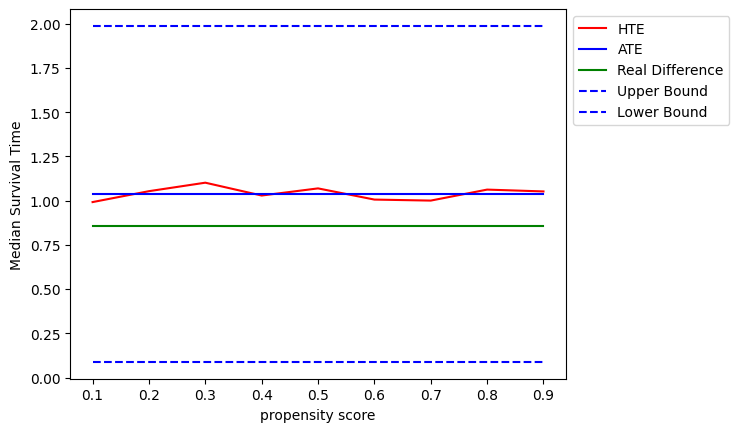}
\caption{ATE and HTE from Naive Plug-in NN-DR Loss model}\label{fig5}
\begin{figurenotes}
    The standard deviation is bootstrapped from 100 loops. 
\end{figurenotes}
\end{figure}

For the NN- DR loss naïve plug-in estimator, since we do not assume the underlying distribution for the survival function, we can not directly estimate the median survival time with the survival probability. Thus, we assume that the survival probabilities estimated from the doubly robust survival loss are at the maximum of the observed time for individuals. With this assumption, obviously, the estimation from the naïve plug-in NN-DR loss estimator will inflate. \\

The naïve plug-in NN-DR loss estimator results can be seen in Figure \ref{fig5}. As we speculated, the estimation of the ATE surpasses the real value at 1.03, and the bootstrapped standard deviation from this model is 0.48. The estimation of the HTE is not linear, with the lowest at the seventh decile at 0.97 and the highest at the third decile at around 1.10. In general, the relationship between the ATE line and the HTE line from the naïve plug-in estimator is the one that best fits our vision. \\

\subsection{Results from Doubly Robust Causal Estimator}
In this subsection, we present the results from the doubly robust causal estimator, as the algorithm demonstrated in Section \ref{sec6}. In short, to calculate the doubly robust causal estimator \( E\bigg[\big[\frac{W_{i} - \hat{\pi}}{\hat{\pi}(1-\hat{\pi})} (Y - \widehat{\mu}_{W_{i}})+ (\widehat{\mu_1} - \widehat{\mu_0})\big] | X = X_{i}\bigg]\), we split the dataset into two subsets, with one subset to predict the mean of \(\frac{W_{i} - \hat{\pi}}{\hat{\pi}(1-\hat{\pi})} (Y - \widehat{\mu}_{W_{i}})\) and another one to fit the mean of \((\widehat{\mu_1} - \widehat{\mu_0})\). Indeed, the latter part is what we fit with the naïve plug-in estimators. Thus, we only need a model estimating the inverse probability weighting and the predicted outcome from the full-data model (not splitting into treatment and control models).\\
\begin{figure} 
\includegraphics[width = 1\textwidth]{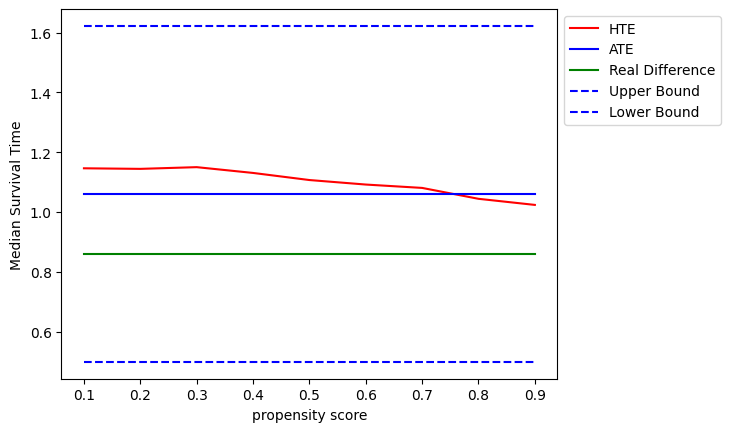}
\caption{ATE and HTE from Doubly robust causal estimator with cox-ph survival}\label{fig6}
\begin{figurenotes}
    The standard deviation is bootstrapped from 100 loops. 
\end{figurenotes}
\end{figure}
We first fit the doubly robust causal model with the Cox-PH model to estimate the survival functions. The results are shown in Figure \ref{fig6}. The estimated median survival time between the treatment group and the control group is 1.06; thus, we can infer that the IPW part upward modifies the estimation of the ATE. The standard deviation for the ATE is 0.29, a little bigger than the standard deviation from the naïve plug-in Cox-PH estimator. The HTE line suggests that the treatment effect still declines as the propensity score increases, but the slope is steeper for the higher propensity score range and smoother in the lower range. \\
 \begin{figure} 
\includegraphics[width = 1\textwidth]{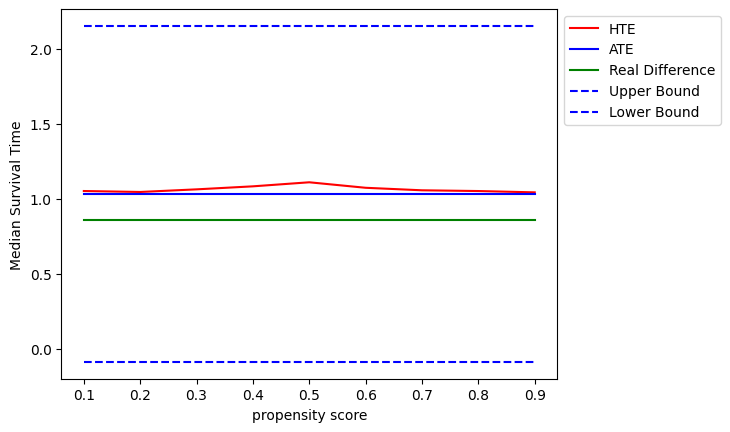}
\caption{ATE and HTE from Doubly Doubly robust estimator}\label{fig7}
\begin{figurenotes}
    The standard deviation is bootstrapped from 100 loops. 
\end{figurenotes}
\end{figure}
Finally, we have the results from the doubly doubly robust estimator- doubly robust estimator for the causal estimation and neural network with a doubly robust loss for the survival function. The calculation of the loss function, the neural network architectures, and the machine learning training settings are the same as in the naive plug-in estimator, and we still assume that the survival probabilities estimated from the doubly robust loss survival functions are the survival probabilities at the maximum observed event time, which may lead to the overestimation. The results are presented in Figure \ref{fig7}. The point estimation of the ATE is 1.03, almost unchanged from the same naive plug-in estimator. Meanwhile, the standard deviation from the doubly robust causal model also increases to 0.57 (compared with 0.48 in the naive one). Moreover, as can be inferred from the Figure, the HTE line almost perfectly fits the ATE line (with point estimators ranging from 1.02 to 1.10). \\

Based on the results, we can conclude that the doubly doubly robust estimator has the most accurate estimation of the ATE among the models we examined here. The doubly robust Cox-PH model yields very similar results. However, we need to remind the readers that this is because the simulated data do not violate the proportional hazard assumption, and the Cox-PH models could perform well with the underlying distributions. In empirical data, the underlying distribution and the data-generating process might be more flexible, and Cox-PH models may not be applicable. However, a relative disadvantage of the doubly doubly robust estimator is that, because it assumes a flexible distribution, we could not directly yield the survival time estimation, and further assumptions are required to calculate the median survival time. In our model, we assume the survival probabilities estimated from the doubly robust loss survival functions are the survival probabilities at the maximum observed event time, which is a source of the overestimation of the effects. 

\section{Conclusion and Further Discussions}
In this chapter, we introduce a doubly doubly robust estimator to estimate the average and heterogeneous treatment effects for the left-truncated-right-censored survival data: we conduct doubly robust estimation both for the causal estimand and for the survival functions and combine them. It is a regular and asymptotically linear and efficient estimator used to identify causal effects. \\

Some readers may have noticed that in this chapter, we assume that the treatment is an instantaneous variable and didn't discuss the scenario of the time-varying treatment. In fact, we believe that for time-varying treatment, calculation on the average and heterogeneous treatment effect is not a statistical/methodological problem; instead, it is a theoretical problem as we need to use a theoretical framework to identify what the "treatment effect" stands for. For instance, suppose we would like to discuss the effectiveness of a drug for treating advanced-stage cancer. We may administer the medication to the patient in the -1st, 0th, 1st, 2nd... weeks as the cancer cells spread from the primary organ to other organs. Therefore, we may have the average treatment effect of the drug for the -1st, 0th, 1st, 2nd... weeks (multiple ATEs), as we can identify the treatment and the control groups in the -1st, 0th, 1st, 2nd... weeks (the treatment group was given the drug in the corresponding week while the control group was given the placebo). We may also average the ATEs for the -1st, 0th, 1st, 2nd... weeks and claim this is the average treatment effect of the drug. Whether the average of the ATEs from the several weeks is valid requires cancer experts to judge. We will see a very similar problem in the next chapter when we apply the model to empirical studies of the causal effect of widowhood on mortality. 

\bibliographystyle{chicago}
\bibliography{try.bib}

\end{document}